\documentstyle[twocolumn,prl,aps]{revtex}
\renewcommand{\widetext}{\end{multicols} \global\columnwidth42.5pc}  

\input{epsf.tex}
\def\inseps#1#2{\def\epsfsize##1##2{#2##1} \centerline{\epsfbox{#1}}}

\begin{document}
\twocolumn[\hsize\textwidth\columnwidth\hsize\csname @twocolumnfalse\endcsname

\draft

\title{Incompressible paired Hall state, stripe order and the
composite fermion liquid phase in half-filled Landau levels}

\author{E. H. Rezayi$^a$ and F. D. M. Haldane$^b$}
\address{
$^a$Department of Physics, California State University, Los Angeles, 
California 90032\\
$^b$ Department of Physics, Princeton University, Princeton, New Jersey 08544}

\date{June 1999; revised March 24, 2000}

\maketitle
\begin{abstract}
We consider the lowest two Landau levels at
half filling. In the higher Landau level ($\nu =5/2$), we find a
first-order phase transition separating a compressible
striped phase from a paired quantum Hall state, which is
identified as the  
Moore-Read state.
The critical point is very
near the Coulomb potential and the transition can be driven by
increasing the
width of the electron layer. We find
a much weaker transition (either
second-order or a crossover) from pairing
to the composite fermion Fermi
liquid behavior.  A very similar picture is obtained for the lowest Landau
level but the transition point is not near the Coulomb potential.
\end{abstract}

\pacs{73.20.Dx, 73.40.Kp, 73.50.Jt}
]
A two-dimensional electron gas in an intense perpendicular 
magnetic field displays 
a host of collective ground states.  The underlying reason is the
formation of two-dimensional Landau levels in which the kinetic energy
is completely quenched.  In the macroscopically-degenerate Hilbert
space of a given Landau level, 
only the Coulomb potential remains, making the system 
strongly interacting.  The fractional quantum Hall effect\cite{books},
at rational fillings of the Landau levels,
is one instance 
of such a ground state (GS).  Other examples occur at
half integral fillings of Landau levels.
In the lowest Landau level, $\rho_{xx}$ shows a shallow minimum and no
plateau\cite{willet} in $\rho_{xy}$.  This behavior has been associated with
a compressible 
Fermi-liquid-like state\cite{hlr} of composite fermions\cite{jain} (CF). 
In sharp contrast, a plateau in $\rho_{xy}$
and activated
$\rho_{xx}$ has been observed at filling factor $\nu=5/2$\cite{fivehalf},  
indicative of an incompressible quantum Hall state. 
Above the second Landau level, for $\nu=9/2,11/2,13/2,$ the transport
is highly anisotropic\cite{lilly,du,shayegan},
suggesting the GS is a 
compressible charge density wave
(CDW) stripe state \cite{koulokov,moessner,RHY}.

Some years ago we proposed\cite{HR} a spin-singlet wavefunction 
$\Psi_{\rm HR}$ for the
5/2 effect based on the idea of  electron
pairing\cite{halperin}.
Moore and Read\cite{MR} (MR), building on the analogy of this state to
Bardeen-Cooper-Schrieffer pairing of CF's, 
proposed a similar spin-polarized pairing wavefunction $\Psi_{\rm MR}$: 
\begin{eqnarray}
\Psi_{\mbox{HR}}(\{z_i,\alpha_i,\beta_i\})
&=& \mathop{\rm Pf}_{i,j}
\left [{\alpha_i\beta_j-\beta_i\alpha_j\over (z_i-z_j)^2}\right  ]
\Psi_L^{(\nu = 1/2)} , \label{hr}\\
\Psi_{\mbox{MR}}(\{z_i,\alpha_i,\beta_i\}) &=&
\mathop{\rm Pf}_{i,j}\left [{\alpha_i\alpha_j\over z_i-z_j}\right ]
\Psi_L^{(\nu = 1/2)} , 
\end{eqnarray}
where $\alpha$ and $\beta$ are spinor coordinates for up and down spins, 
$\mathop{\rm Pf}[A]$ is the Pfaffian of an antisymmetric matrix $A\cite{form}$,
and $\Psi_L^{(\nu = 1/2)}$ is the Laughlin state (for bosons).

Subsequently, Greiter, Wen and Wilczek (GWW)\cite{gww} suggested that the MR
state may be a possible candidate for the 5/2 effect.  Recent numerical 
calculations by Morf\cite{morf} show the polarized state to
have a lower energy than spin-singlet states even without Zeeman
energy.  Yet,  these studies have not established what the true
nature of the 5/2 state is.
In this paper we present evidence which suggests that the $\nu=5/2$ effect
indeed derives from a paired state which is closely 
related to the MR polarized state or, more precisely, to the state obtained
by particle-hole (PH) symmetrization of the MR state. We also show why 
the transport may not be quantized\cite{tilt0} and may become anisotropic 
upon tilting the field, as observed\cite{tilt1,tilt2}.  
We find a first-order phase transition from   a 
striped phase to a strongly-paired state,
after which the system evolves into a Fermi-liquid-like state, either 
by a continuous crossover to a weakly-paired state,
or a second-order transition to a gapless state (our 
calculations cannot distinguish
these possibilities).

Our conclusions are 
based on numerical studies  for up to 16 electrons in
two geometries: sphere and 
torus.  The torus is particularly convenient for investigating 
the nature of the ground state at $\nu=1/2$.  All three states of 
interest---
composite fermion Fermi surface, pairing  and CDW---
are realized at flux $N_\phi=2N$ (in units of flux quanta). This
avoids a problem on the sphere, where, for a given $N$,
different $\nu$ = 1/2 states
occur at slightly different $N_{\Phi}$.
We only consider states within a given Landau level and discard the
kinetic energy.  The Hamiltonian is:
\begin{equation}
H=\sum_{m=0}^{\infty}V_m
{2 \over N_{\Phi} } \sum_{\bf q}
e^{-\mbox{$\case{1}{2}$}q^2}L_m(q^2)
\sum_{i<j}  e^{i{\bf q}\cdot(
{\bf R}_i-{\bf R}_j)},
\label{ham}
\end{equation}
where ${\bf R}_i$ is the guiding center\cite{haldane}
coordinate of the $i$'th electron, $L_m(x)$'s are the 
Laguerre polynomials, and $V_m$ is the energy of a pair of electrons in 
a state of relative angular momentum $m$.
These are the pseudo-potential parameters\cite{haldane,foot1}.  
The magnetic length is set to 1.
Unless otherwise specified the data presented here is for ten fully-polarized
electrons in a hexagonal unit cell. 

The Fermi-liquid  state is well-described by a Fermi sea of 
composite fermions\cite{rr,group}, which on the torus is\cite{group}:
\begin{equation}
|\Psi_{CF}(\{{\bf k_i}\}\rangle=
\det_{i,j}[ \exp(i {\bf k}_i\cdot {\bf R}_j)] |\Psi_L^{(\nu = 1/2)}\rangle,
\end{equation}
where the
$\{ {\bf k}_i\}$ 
are distinct (and belong to the usual set of wavevectors
allowed by the PBC's) and are
clustered together to form a 
filled ``Fermi sea'' centered on ${\bf k}_{\rm av}=\sum {\bf k}_i/N$.
The total momentum quantum number ${\bf K}$ \cite{halpbc} 
is determined
by the value of ${\bf k}_{\rm av}$  {\it relative to the set of allowed
${\bf k}$'s} (the CF state is essentially left invariant 
by a uniform  ``boost'' 
$\{ {\bf k}_i \}$
$\rightarrow $
$\{ {\bf k}_i +{\bf k}\}$), and
takes one of $N^2$ distinct values\cite{halpbc}.
There are four distinct values of 
${\bf k}_{\rm av}$ which are invariant under $180^{\circ}$-rotation:  
${\bf k}_{\rm av}$ = 0, 
and  ${\bf k}_{\rm av}$ halfway between allowed ${\bf k}$-vectors (three distinct
values which correspond to  the three distinct values of ${\bf K}$ for the
MR state on the torus).

The $\nu$ = 1/2 spin-polarized electron eigenstates of  
(\ref{ham}) have particle-hole (PH)
symmetry\cite{foot2};
the CF state is {\it almost} 
(99.935\%) PH-symmetric 
and also has a large projection (99.25\%)  on
the exactly PH-symmetric  GS of the 
Coulomb potential in the lowest Landau level. 

The periodic MR states\cite{gww} can be obtained 
as the zero-energy ground states of a 3-body short range 
potential\cite{gww},  the corrected form of which is
\begin{eqnarray*}
H_3 &=& -\sum_{i<j<k} {\cal S}_{i,j,k}\{\nabla_i^4\nabla_j^2\}
\delta^2({\bf r}_i-{\bf r}_j)\delta^2({\bf r}_j-{\bf r}_k),
\end{eqnarray*}
where ${\cal S}_{i,j,k}$ is a symmetrizer.
Note that (in contrast to (\ref{ham})) $H_3$ has no PH symmetry
and the MR state does {\it not} possess definite parity under
PH transformations.
\begin{figure} 
\inseps{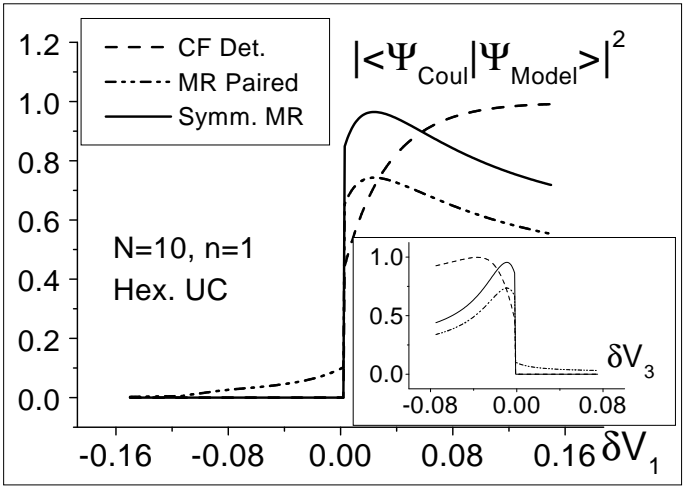}{0.35}
\caption{The projection of the exact GS of the Coulomb
interaction in the $n=1$ Landau level, plus an extra short-range
pseudopotential $\delta V_1$ ($\delta V_3$ in the inset),
on the CF, MR, and PH-symmetrized MR model states. 
The GS PH parity changes at a level crossing near
$\delta V$ = 0.
} 
\label{fig:1}
\end{figure}
The nature of the ground state of (\ref{ham}) depends on the relative strengths
of the  pseudo-potentials,  in particular 
$V_1$ and $V_3$ (even-$m$ pseudo-potentials do not affect polarized 
states).  
Figure  \ref{fig:1} and Fig. \ref{fig:4} (below) show the projection of the CF and MR  
states on the
exact GS in two different PBC geometries, 
as $V_1$ and $V_3$ are varied relative to their Coulomb
values in the first excited Landau level ($n$=1).
Varying $V_3$ alone (the inset of Fig. \ref{fig:1}) or varying both $V_1$ and $V_3$, yields similar 
results, though $\delta V_1$ has an opposite effect
to $\delta V_3$.  A study
using spherical geometry \cite{morf} also 
identifies the phase at large $\delta V_1$ with the CF liquid. 
\begin{figure}
\inseps{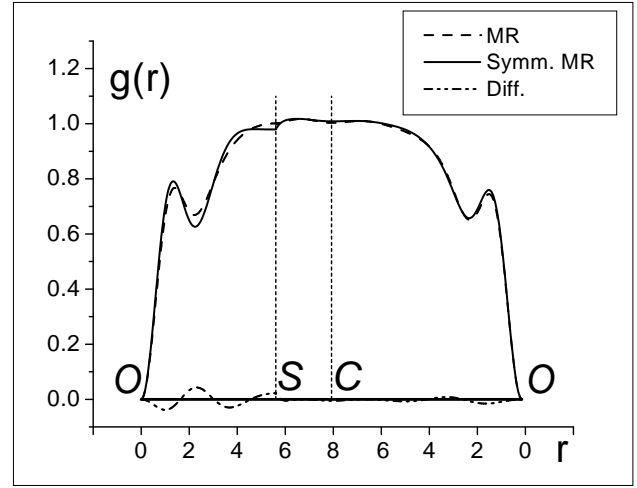}{0.346}
\caption{The real-space pair-correlation function for the MR state and its PH-symmetrized
counterpart, evaluated in the second ($n$ = 1) Landau level);   
their difference is also shown.
$g({\bf r})$ in a square unit cell is shown along a path from
the origin $O$  to the midpoint of a side $S$, to a corner $C$,
and back to $O$.}
\label{fig:2}
\end{figure}
\noindent
A first-order phase transition from a compressible state to an incompressible
paired state
is clearly seen.  
The transition is very close to the Coulomb 
value 
($\delta V_1$ = $\delta V_3$ = 0).  We obtain similar results in the
lowest Landau level, except that the transition point occurs at $\delta V_1= 
-0.092,\ \ \delta V_3=0$ and at $\delta V_1=0,\ \ \delta V_3 = 0.048$.  Details of these studies will
be given elsewhere.  For both Landau levels, we only observe the strongly
paired 
state in a narrow window.
The projection of the MR state on the 
exact ground state does not exceed 73\% in 
this region.  
However,  if the MR state is first PH-symmetrized,
this projection becomes 97\%.
The two-particle correlation
function $g({\bf r})$
of the states before and after symmetrization is shown 
in Fig. \ref{fig:2}.  
The paired character of the 
MR state is essentially unaltered (Fig. \ref{fig:2} shows
that each electron has one particularly close partner); 
the near isotropy of $g({\bf r})$ is 
characteristic of the incompressible states, and should improve 
with increasing system size.

\begin{figure}
\inseps{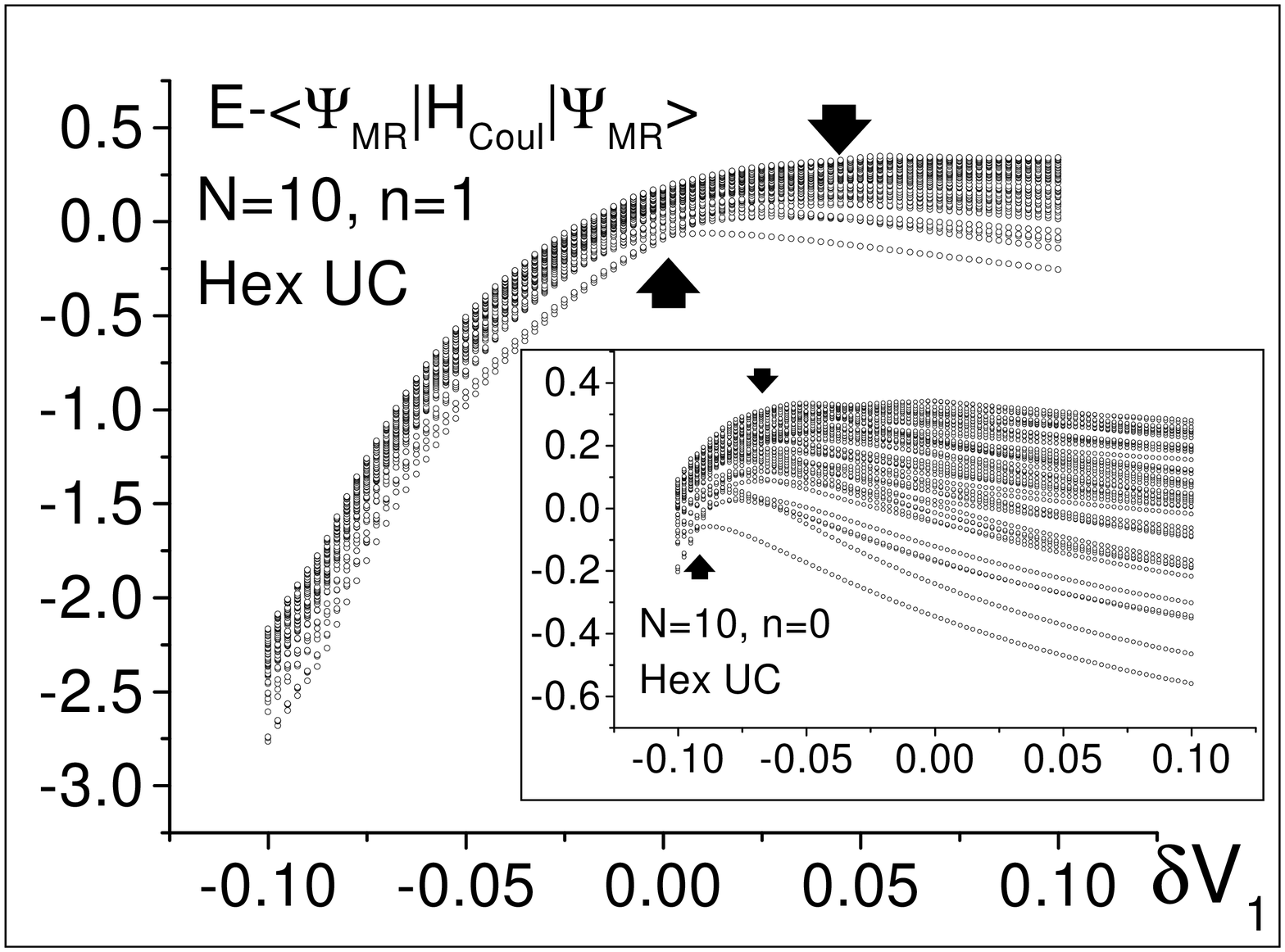}{0.35}
\caption{The low-lying spectrum (relative to the variational energy of the 
MR state)plotted vs. $\delta V_1$ for the $n=1$; the inset
shows this for the $n$ = 0 LL.  The Coulomb point is
$\delta V_1 = 0$. The
energies are scaled by the bandwidth of the two-particle system. The region 
between the arrows is the strong pairing regime.} 
\label{fig:3}
\end{figure}

An interesting feature in Fig. \ref{fig:1} 
is the absence of any obvious sharp transition from the
paired state to the compressible Fermi-liquid-like CF state
as $V_1$ is increased further.  This is also seen in the 
excitation spectrum.  Figure \ref{fig:3} shows the low-lying excitation spectrum
as a function of $V_1$.  Again, there is only one 
first-order level crossing transition (shown by up-arrows).  
The levels that cross have the same translational and 
$180^{\circ}$-rotation symmetry but belong to opposite 
parities under PH transformation.  The MR state has a 
finite overlap with the exact GS on {\it both} sides 
of the transition as it has components with both
PH symmetries.
As $\delta V_1$ increases further,
the excitation spectrum  
gradually evolves from having a clear gap to the compressible 
CF Fermi-liquid-like spectrum\cite{rr,group}.  
The crossover 
is approximately at the point where the spectrum begins to change 
at the level crossings of the {\em excited} states (down-arrows).
Similar 
crossover behavior is also seen on the sphere, and in 
those  geometries on the torus where the most-compact
Fermi sea has $180^{\circ}$-rotation symmetry,
so the CF state has the same ${\bf K}$ 
as the MR state.

The hallmark of compressible CF states is the sensitivity of the GS ${\bf K}$ 
to the PBC
geometry. For example, the Fermi surface for 10 electrons with the 
square
PBC does not have $180^{\circ}$-rotation 
symmetry and has a ${\bf K}$  different from  
the MR
state. A sharp transition is seen in this case (Fig. \ref{fig:4}).  
As Figs. \ref{fig:1} and \ref{fig:4} clearly demonstrate, the evolution
to the CF state
is strongly dependent on geometry while the transition to the striped phase
is not.  We believe this rules out a first-order transition to the CF liquid
state.
The picture most
consistent with our studies is that, after the first-order transition 
to the paired state,
the system may {\it always} be paired,
and smoothly crosses over from a strong to
a weak pairing regime as the interaction is varied.  
In the weak pairing regime, such
a system would exhibit CF Fermi-liquid behavior
at energy scales and temperatures above the gap
and paired quantum Hall behavior 
below the gap; finite-size effects in our calculations
will mask a very small gap.   If true, this would
eliminate the infra-red divergences  of
\cite{hlr}.

In agreement with this, we find substantial pairing
character in the lowest Landau level for the Coulomb potential
in both spherical and toroidal geometries.  For example, on the sphere
(with flux $N_\phi=2N-3$) we found that the projections of the MR state 
on the exact ground state 
of the Coulomb potential increases with system size 
(43\%, 52\%, 56\% at $N$=12, 14, 16),
even though the relevant $L=0$ Hilbert space
grows twenty-fold.  This would be consistent with weakly-bound
pairs that are larger than the linear system size at small $N$; 
however,  because we cannot study larger $N$, we are unable to
conclusively exclude the possibility
of a second-order (or even a weakly first-order)
phase transition to a gapless  CF state.

\begin{figure}
\inseps{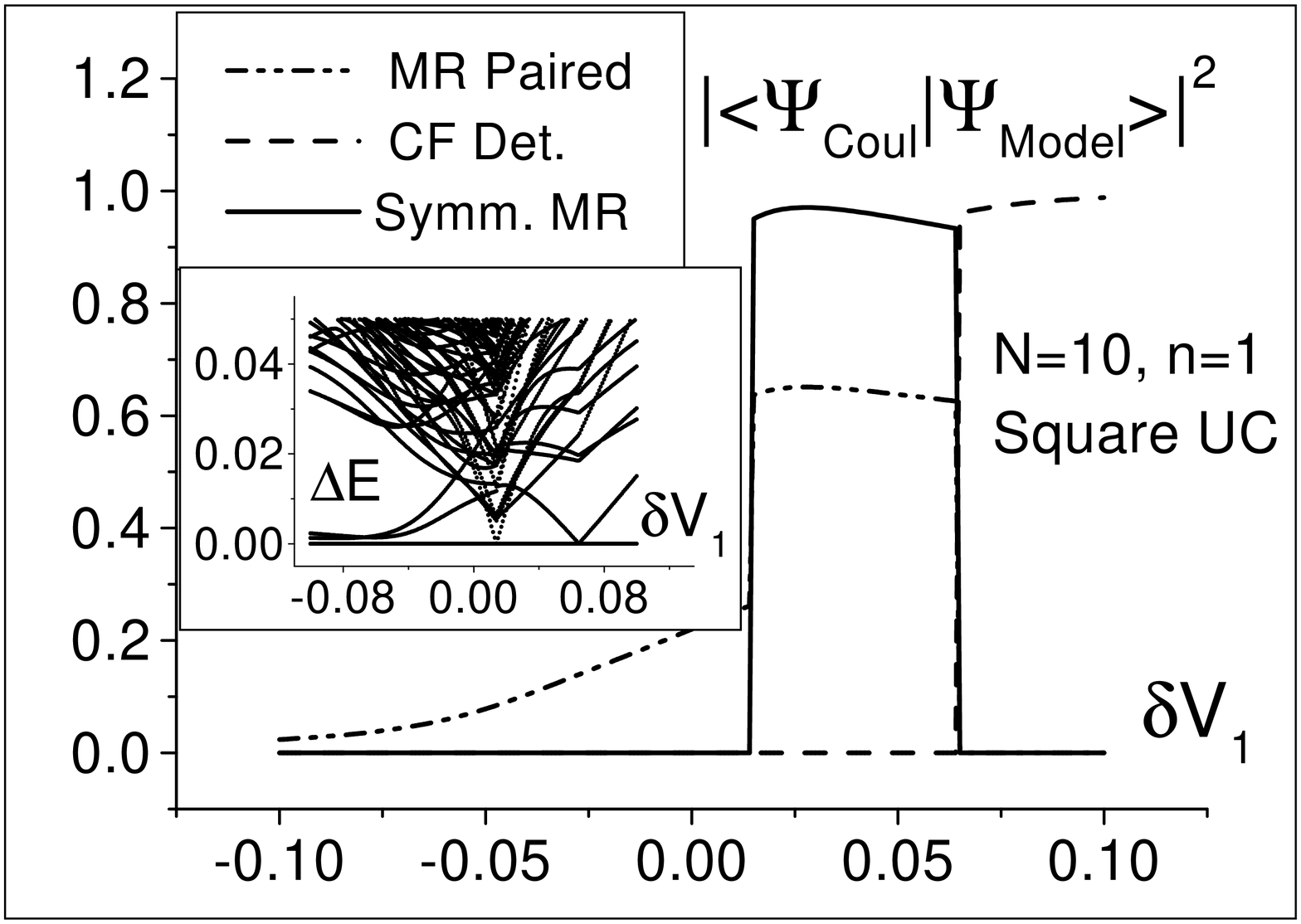}{0.35}
\caption{Same as Fig. \ref{fig:1} but for a square unit cell.  The inset shows the 
excitation spectrum (the GS energy is subtracted) as a function 
of $\delta V_1$.  The transition points are marked by the collapse of the 
gap.  In the striped phase (left portion) one recovers the 
typical degeneracies seen in the $n=2,3$ LL's.} 
\label{fig:4}
\end{figure}

We next turn to the compressible state seen to the left of the transition
in Figs. \ref{fig:1}, \ref{fig:3} and \ref{fig:4}. To show its character more clearly, we reduce $V_1$
by 0.05 (about 10\% of its Coulomb value).  
Here, as in the Fermi-liquid 
state, the GS ${\bf K}$-vector changes with size and geometry indicating the
state is compressible.  We now consider 12 electrons in 
a rectangular unit cell
and tune the  aspect ratio to 0.5.  We find  two strong peaks in
the static guiding center structure function $S_0({\bf q})$
and in the charge susceptibility
$\chi({\bf q})$ with ordering wavevector ${\bf q^*}=(1.1,0)$ which constitute
the signature of the CDW stripe ordering\cite{RHY}.  
This system forms three stripes
and the weight of the single Slater determinant state 
with the occupation pattern
000011110000111100001111 is about 58\%.  Edge fluctuations of stripes
seem to be
stronger here than in the higher Landau levels;
$V_1$ has to be somewhat reduced below the transition for  
the characteristic degeneracies of the broken symmetry phase to be well
developed (inset of Fig. \ref{fig:4}). 

\begin{figure}
\inseps{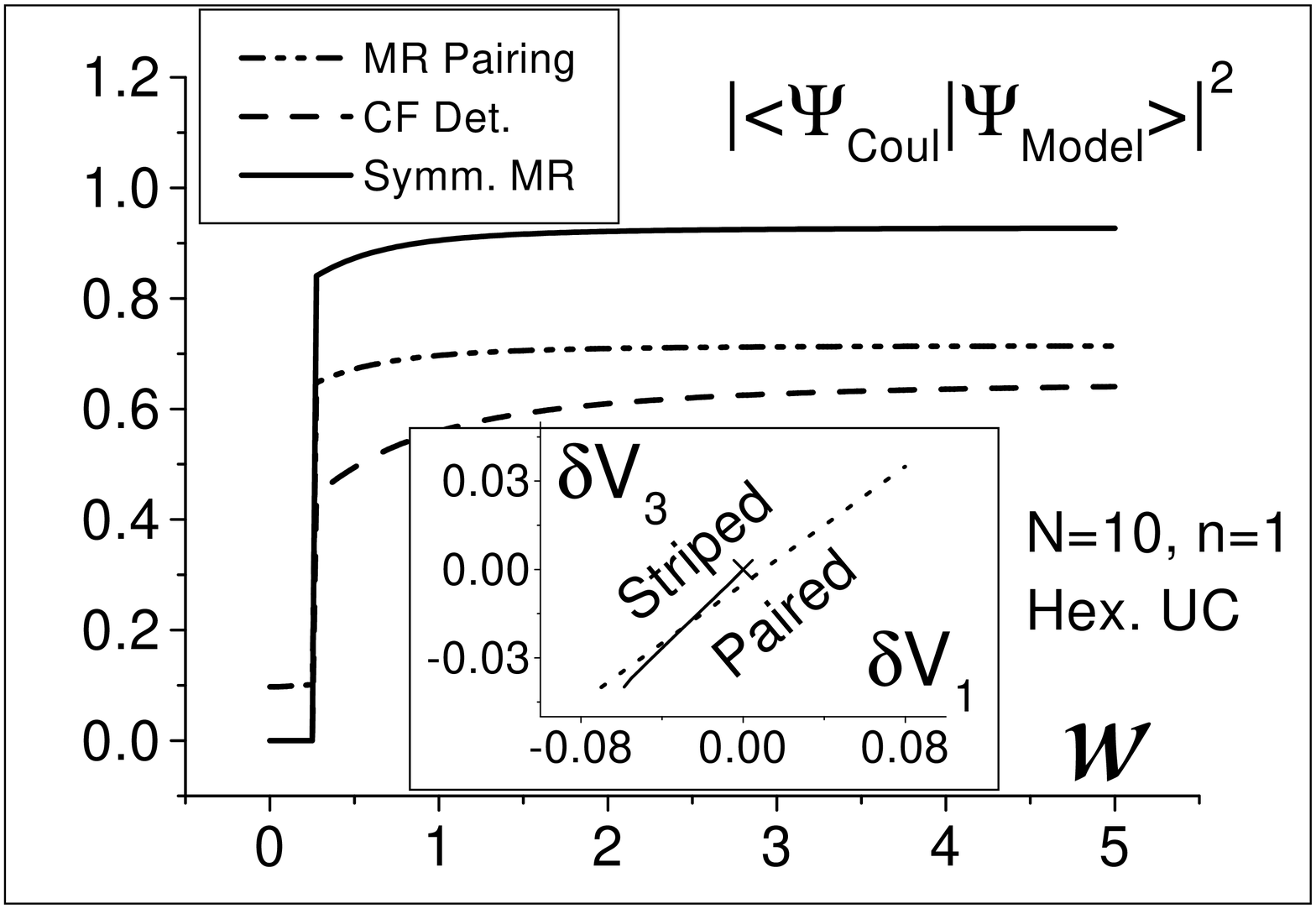}{0.35}
\caption{The overlap squared of the two model states as the layer width $w$ is
varied in the $n=1$ Landau level. The inset shows the boundary between striped 
and paired phases and how layer thickness changes $\delta V_1$ and $\delta V_3$
as $w$ is varied from 0 (at the cross) to 1. The system crosses the phase 
boundary at $w=0.3$ along the solid line.}
\label{fig:5}
\end{figure}

We believe that the {\em proximity} of the critical point to the Coulomb 
potential 
is the principal reason for the disappearance of the paired Hall state
upon tilting the field\cite{frsphere}. 
One effect of the tilted field is to compress the 2-D 
layer\cite{gww,jim}.  Indeed,
we have found that varying 
the layer width
drives this transition (as suggested by GWW\cite{gww})
in most of the pbc geometries that we have studied.
The critical width varied from 0.23 to 2.4 in these systems.  
Fig. \ref{fig:5} shows the overlap (squared) as a function of the
layer width in the $n=1$ Landau level. We have used the Fang-Howard model for
layer profile (with $w=2b$)\cite{haldane,prange}.  
In the lowest Landau level, the GS of the Coulomb potential is well in the
CF regime. 
The projection on the MR state
increases from 54\% for a thin layer to 64\% for very
wide layers (83\% on the PH-symmetrized MR state).
For both Landau levels, increasing the layer width increases
the pairing correlations, as seen also in Monte-Carlo calculations\cite{jain2}.

We acknowledge useful discussions   with R. Morf,  J.
Eisenstein, and especially N. Read.  This work was 
supported by NSF DMR-9420560 (E.H.R.) and 
DMR-9809483 (F.D.M.H.). We thank ITP-UCSB for their  hospitality during the
``Disorder and Interactions in Quantum Hall and Mesoscopic Systems" workshop
supported by NSF-PHY94-07194.

{\it Note added}---A realistic potential taking into account 
finite layer width, screening by filled Landau levels and 
tilted field effects, including mixing of subband levels,
(modeled for Eisenstein's experimental samples and supplied to us by 
Girvin, Jungwirth and MacDonald),
confirms that (a) the paired state at $\nu=5/2$ and zero tilt 
is indeed described by the symmetrized MR state (with 98\% weight) and (b)
tilting drives the system into a stripe phase.  Details of these studies will
be given elsewhere.


\begin{references}
\bibitem{foot1} For the Coulomb potential in $n=0$ LL these are 
$V_m^0=\Gamma(m+1/2)/(2\Gamma(m+1))$, and 
for $n=1$ they are $V_m^1=(m-3/8)(m-11/8){\Gamma(m-3/2)/(2\Gamma(m+1))}$.
\bibitem{frsphere} Morf has come to the same conclusion from studies on
the sphere where the transition is not as sharp and $V_1$
was reduced by 20\% from the ``maximum gap'' point.
\end{references}
\end{document}